\documentclass[twocolumn]{aastex631}
\usepackage{amsmath,amssymb, amsthm,amstext}
\usepackage{natbib}
\usepackage{graphicx, float, subfigure}
\usepackage{color}
\usepackage{array, enumerate}
\usepackage{bm}
\usepackage{multirow}
\usepackage{braket}
\usepackage{txfonts}
\usepackage{physics}
\usepackage[normalem]{ulem}
\usepackage{appendix}

\def\be{\begin{equation}}
\def\ee{\end{equation}}

\begin{document}

\title{No evidence for a supermassive black hole binary in GSN 069}

\author[0009-0001-2065-9641]{Yuhe Zeng}
\email{zengyuhe@sjtu.edu.cn}

\author[0000-0001-9608-009X]{Zhen Pan}
\email{zhpan@sjtu.edu.cn}
\affiliation{Tsung-Dao Lee Institute, Shanghai Jiao-Tong University, Shanghai, 520 Shengrong Road, 201210, People’s Republic of China}
\affiliation{School of Physics \& Astronomy, Shanghai Jiao-Tong University, Shanghai, 800 Dongchuan Road, 200240, People’s Republic of China}

\author{Bin Liu}
\affiliation{Institute for Astronomy, School of Physics, Zhejiang University, Hangzhou310027, People’s Republic of China}
\affiliation{Center for Cosmology and Computational Astrophysics, Institute for Advanced Study in Physics, Zhejiang University, Hangzhou 310027, People’s Republic of China}

\author{Cong Zhou}
\affiliation{Department of Astronomy, University of Science and Technology of China, Hefei 230026, People’s Republic of China}

\begin{abstract}
Quasi-periodic eruptions (QPEs) are recurrent soft X-ray flares from galactic nuclei and provide a new time-domain probe of stellar-mass objects (SMOs) orbiting supermassive black holes (SMBHs). In an extreme-mass-ratio inspiral (EMRI) system interacting with an accretion disk, QPEs are produced when the SMO repeatedly crosses an accretion disk, so that the eruption times trace the orbital motion of the EMRI. We investigate whether such timing information can be used to probe a more distant SMBH companion. We develop two complementary diagnostics: (1) the motion of the EMRI host SMBH around the SMBH-binary (SMBHB) center of mass induces a light-travel-time modulation in the observed QPE arrival times, specifically an \emph{in-phase} modulation in arrival times of even and odd eruptions; 
(2) if the QPE source contains a surviving stellar orbiter, the external SMBH must not drive the SMO into tidal disruption through eccentricity excitation by the von Zeipel--Lidov--Kozai (ZLK) mechanism. Using GSN 069 as an example, 
we find \emph{no} in-phase modulation in the QPE timing (i.e., no evidence for an SMBHB) and constrain the excluded parameter space of the companion SMBH.
These results demonstrate that QPE timing and stellar survival offer complementary routes for constraining otherwise hidden SMBH companions in nearby galactic nuclei.
\end{abstract}

\section{Introduction} \label{sec:intro}
\begin{figure*}[!ht]
\centering
\includegraphics[width=0.9\textwidth]{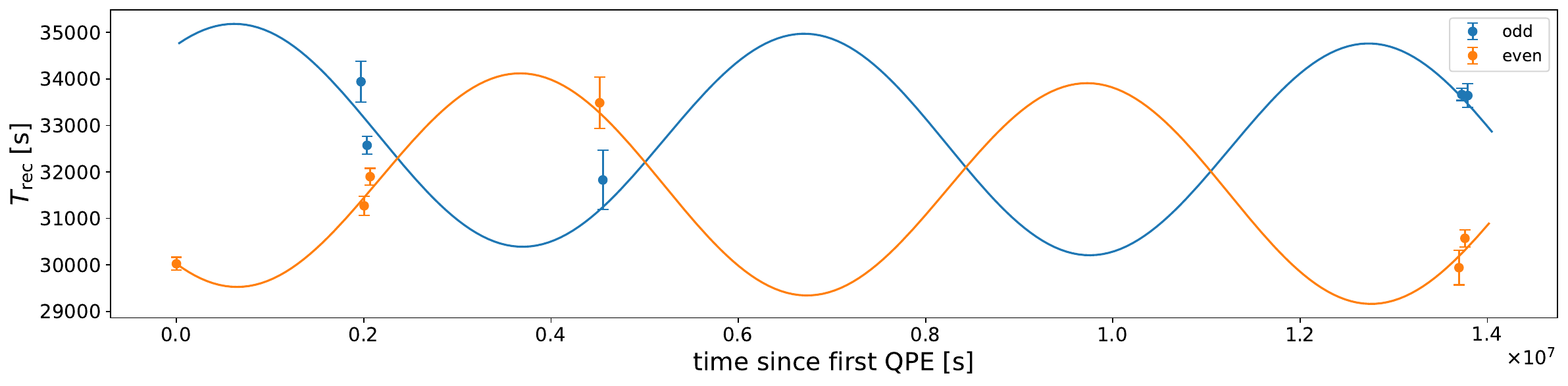}
\caption{Time intervals between adjacent eruptions of GSN 069~\citep{Miniutti2019}. There are clearly two anti-phase branches, denoted as $T_{\rm even}(t)$ and $T_{\rm odd}(t)$, where both of them vary with a period $\approx 76$ d, while the sum $T_{\rm even}(t)+T_{\rm odd}(t)\approx 65$ ks~\citep{10.1093/mnras/staf1580,Zhou_2025}.
No evidence for an in-phase modulation in even and odd eruptions is revealed.}
\label{fig:GSN Padj}
\end{figure*}

Quasi-periodic eruptions (QPEs) are large-amplitude, recurrent soft X-ray flares from galactic nuclei, with recurrence times ranging from hours to days~\citep{Sun_2013, Miniutti2019,Giustini2020,Arcodia2021,Nicholl2024,Chakraborty_2025,HernandezGarcia2025,Arcodia_2025,Baldini_2026}, and host black hole masses typically inferred to be below $\sim10^7\ M_\odot$~\citep{Wevers2022,Arcodia2024,HernandezGarcia2025}. The discovery source GSN 069~\citep{Miniutti2019} shows one of the clearest timing patterns: its recurrence intervals alternate between long and short branches, and the two branches vary approximately in anti-phase over a timescale of tens of days (see Fig.~\ref{fig:GSN Padj}). Similar nuclear transients have now been found in a growing number of sources, including systems discovered  in tidal disruption events (TDEs)~\citep{Nicholl2024,Chakraborty_2025,Zhu:2025rqb,Wu:2026pkl}. 
GSN 069 exhibited a long-term soft X-ray decay before the appearance of QPEs, suggestive of a previous nuclear accretion episode~\citep{Shu_2018,Miniutti2019,Miniutti2023,Miniutti_2023}. 
The TDE-QPE association has an important physical implication that some QPEs occur in galactic nuclei where a compact accretion disk has been newly formed.
From the perspective of the host galaxies, QPE and TDE hosts show similar morphologies~\citep{Gilbert_2025}, while extended emission line regions (EELRs) have been detected in several QPE hosts~\citep{Wevers_2024}, including the joint TDE--QPE host AT2019qiz~\citep{Xiong_2025}. A large fraction of TDE-QPE association pairs 
have been found with bright TDE infrared echoes \citep{Wu_2025, Baldini_2026}.



Several classes of models have been proposed for QPEs, including accretion-flow instabilities~\citep{Sniegowska_2020,Raj_2021,Pan_2022,Pan_2023,Sniegowsk_2023,10.1093/mnras/stad1894}, repeated partial disruptions and mass transfer~\citep{MacLeod_2013,King_2020,10.1093/mnras/stac1641,10.1093/mnras/stad2203}, and collisions between an orbiting object and an accretion disk~\citep{Dai_2010,Xian_2021,Linial_2023,Franchiniqpestardisk_2023,10.1093/mnras/stad2616,PhysRevD.109.103031,PhysRevD.110.083019,Yao_2025,liu2026quasiperiodiceruptionsstellarmassblack,Jankovic_2026, Huangshunquan_2026,Chen:2026fdv}. 
In this work, we focus on the EMRI+disk interpretation, where repeated disk crossings naturally produce a regular eruption clock, generally with two eruptions per orbital period, while a mildly eccentric orbit with apsidal precession  accounts for the alternating long-short recurrence pattern observed in GSN 069: the recurrence times alternate between $T_{\rm long}$ and $T_{\rm short}$, both varying with time, while $T_{\rm long}(t)+T_{\rm short}(t)$ remains approximately a constant~\citep{Miniutti2019,PhysRevD.109.103031,PhysRevD.110.083019}. The observed association of some QPEs with TDEs also provides a natural origin for the accretion disk~\citep{Linial_2023,Nicholl2024,Chakraborty_2025,Zhu:2025rqb,Wu:2026pkl}.

In the EMRI+disk framework, the QPE timing encodes rich information about the EMRIs, and has been used for measuring the SMBH mass and the EMRI orbital parameters~\citep{Xian_2021,PhysRevD.109.103031,PhysRevD.110.083019,10.1093/mnras/staf1580,Zhou_2025,Chakraborty_2025timing}. The EMRI orbital parameters, particularly the orbital semi-major axis and the eccentricity, are sensitive probes of their formation history ~\citep{PhysRevD.109.103031,PhysRevD.110.083019,Jiang_2025,allievi2026twobodyrelaxationemritdedisk,3gvb-rv11,rom2026dynamicsnuclearstellarclusters,witzany2026multimessengerprospectsquasiperiodiceruptions}.
As summarized by \cite{Jiang_2025}, the majority of QPE EMRIs are of low eccentricities, which are consistent with \emph{Wet} EMRIs~\citep{PhysRevD.103.103018,PhysRevD.104.063007, Zeng:2026ydj}
formed in previous AGN disks. 
If a significant fraction of QPEs are associated with compact-object EMRIs, their inferred occurrence rates may also inform expectations for low-frequency gravitational wave (GW) sources and stochastic backgrounds in the LISA band~\citep{Arcodia2024_LISA_detection,black2026extrememassratioinspirals}.



In this paper, we aim to investigate whether QPEs can be used for constraining a more distant SMBH companion.
SMBH binaries are expected outcomes of galaxy mergers, but sub-parsec binaries remain difficult to identify electromagnetically because their angular separations are generally unresolved and their accretion signatures can be ambiguous~\citep{DEROSA2019101525,10.1093/mnras/stab3713}. A QPE source embedded in a wider SMBH-binary (SMBHB) offers a different opportunity. The external SMBH affects the observed QPEs in two conceptually distinct ways. First, the central SMBH hosting the EMRI moves around the SMBHB center of mass (C.O.M.), introducing an additional light-travel-time (LTT) modulation in the QPE arrival times. This is a direct timing effect. Second, the external SMBH, acting as a tertiary in the hierarchical triple, gravitationally perturbs the EMRI orbit. If the SMO is a normal star, oscillations driven by the von Zeipel–Lidov–Kozai (ZLK) mechanism can excite the inner orbit to high eccentricity and eventually lead to tidal disruption of the star; the survival of a stellar QPE orbiter therefore provides an indirect dynamical constraint on the external companion~\citep{1962AJ.....67..591K,LIDOV1962719,annurev:/content/journals/10.1146/annurev-astro-081915-023315,10.1093/mnras/stu2396}.

We develop both diagnostics using GSN 069 as an example. In Section~\ref{sec:review}, we first review the QPE timing method for measuring the SMBH mass from the alternating long-short recurrence pattern. In Section~\ref{sec:modulation}, we introduce the SMBHB-induced LTT modulation and incorporate it into the QPE timing model. In Section~\ref{sec:indirect probe}, we compute survival maps for a star in the presence of the tertiary SMBH, incorporating single-averaged (SA) perturbations~\citep{Liu_2018} and first-post-Newtonian (1PN) relativistic apsidal precession. 
We summarize the implications for using QPEs as probes of hidden SMBH companions in Section~\ref{sec:summary}. Throughout this work, we use geometrized units with $G=c=1$ unless otherwise stated.

\section{Review of the QPE timing method}\label{sec:review}

Using GSN 069 as an example, we adopt the four observing epochs: XMM3 (ObsID: 0823680101), XMM4 (ObsID: 0831790701), Chandra (ObsID: 22096) and XMM5 (ObsID: 0851180401). The well-known alternating long-short pattern  in recurrence times $T_{\rm rec}(t)$ is shown in Fig.~\ref{fig:GSN Padj}: there are clearly two \emph{anti-phase} branches in time intervals between adjacent eruptions,\footnote{In the recent literature, there are claims of an \emph{in-phase} modulation in recurrence times of even and odd eruptions of GSN 069, which has been interpreted as 
evidence for an SMBHB. These claims are likely false alarms caused by mismatched cycle number assignment in their O-C analyses  as shown by~\cite{zhou2026noteqpetimingfalse} and in Appendix~\ref{appendix:O-C GSN 069}.} 
denoted as $T_{\rm even}(t)$ and $T_{\rm odd}(t)$, where both of them vary with a period $\approx 76$ d and an amplitude $\delta T_{\rm rec} \approx 2$ ks, while the sum remains approximately a constant, $T_{\rm even}(t)+T_{\rm odd}(t)\approx 65$ ks \citep{PhysRevD.109.103031,Zhou_2025}. No evidence for an \emph{in-phase} modulation in even and odd eruptions is revealed.

The QPE timing of GSN 069 has been used for measuring the SMBH mass and the EMRI orbital parameters in the framework of EMRI+disk~\citep{Xian_2021,PhysRevD.109.103031,10.1093/mnras/staf1580}. 
In the EMRI+disk model, two eruptions are produced per orbital period~\citep{liu2026quasiperiodiceruptionsstellarmassblack,Huang_2025,Jankovic_2026,Huangshunquan_2026}, 
a non-zero orbital eccentricity leads to the alternating long-short pattern~\citep{Linial_2023,pasham2024alivestronglykickingstable}, 
and the EMRI apsidal precession imprints an \emph{anti-phase} modulation in even and odd eruptions as shown in Fig.~\ref{fig:GSN Padj}. As a result, one can infer EMRI orbital information 
\begin{equation}
\begin{aligned}
T_{\rm obt} & \approx 65\ {\rm ks}\ , \\ 
    T_{\rm aps} & \approx 76\ {\rm d} \ , \\ 
     e &\approx \frac{\pi}{2} \frac{\delta T_{\rm rec}}{T_{\rm obt}} \approx 0.04\ , 
    \end{aligned}
\end{equation}
where $e$ is the EMRI orbital eccentricity and $T_{\rm aps}$ is the apsidal precession period.
The apsidal precession period and the orbital period are expressed as 
\be \label{eq:kepler}
\begin{aligned}
    T_{\rm obt} & = 2 \pi \left(\frac{A}{M_\bullet}\right)^{3/2} M_\bullet\ , \\ 
\frac{T_{\rm aps}}{T_{\rm obt}} & =  \frac{p}{3M_\bullet}  = \frac{A(1-e^2)}{3M_\bullet}\ ,
\end{aligned}
\ee
where $A$ and $p$ are the EMRI orbital semi-major axis and semi-latus rectum, respectively. 
It is straightforward to obtain the SMBH mass  as
\be\label{eq:Mbh}
M_\bullet = \frac{1}{2\pi} \left(\frac{1-e^2}{3}\right)^{3/2} \frac{T_{\rm obt}^{5/2}}{T_{\rm aps}^{3/2}} \approx 4\times 10^5\ M_\odot\ ,
\ee 
which is consistent with full Bayesian inference results of GSN 069 timing data in previous studies ~\citep{Xian_2021,PhysRevD.109.103031,PhysRevD.110.083019,10.1093/mnras/staf1580,Zhou_2025}.

\section{Direct probe with QPE timing} \label{sec:modulation}

A framework for inferring EMRI dynamics from QPE timing data has been established in previous studies \citep{PhysRevD.109.103031,PhysRevD.110.083019,10.1093/mnras/staf1580,Zhou_2025}. The observed QPE arrival time $t_{\rm obs}$ is written as 
\begin{equation}
      t_{\rm obs}=t_{\rm crs}+\delta t_{\rm geo}+\delta t_{\rm shap}\ ,
\end{equation}
where $t_{\rm crs}$ is the EMRI disk-crossing time, $\delta t_{\rm geo}$ and $\delta t_{\rm shap}$ are the geometric and Shapiro delays of photons propagating from the emission location to the observer, respectively.  To specify the EMRI trajectory in the Kerr spacetime and the EMRI disk crossing time, 
9 orbital parameters are needed: semilatus rectum $p$, eccentricity $e$, polar turning point $\theta_{\rm min}$, initial phases
in three directions $q_{r,{\rm ini}}, q_{z,{\rm ini}}, q_{\phi,{\rm ini}}$, orbital period $T_{\rm obt}$, the derivative of the orbit period 
$\dot T_{\rm obt}$ due to EMRI-disk collisions, and the dimensionless SMBH spin $a$ (the SMBH mass $M_\bullet$ is a function of $T_{\rm obt}, p$ and $e$).  An additional parameter, $\theta_{\rm obs}$, the polar angle of the observer is required for calculating the delays.

\begin{figure}[!ht]
\centering
\includegraphics[width=0.45\textwidth]{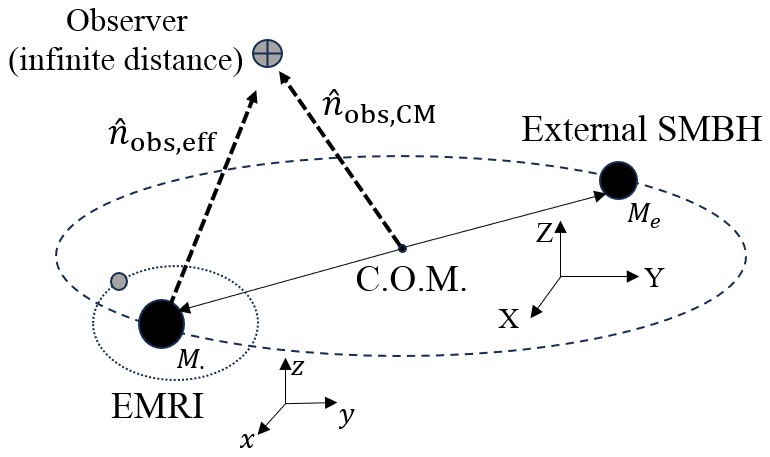}
\caption{Schematic plot of an EMRI system perturbed by an external SMBH $M_{e}$ as the tertiary in the system. The tertiary SMBH $M_e$ forms a wider SMBHB with $M_\bullet$, causing the EMRI center to move around the C.O.M. and thereby inducing an additional LTT modulation in the observed QPE arrival times. The EMRI geometry is described in the coordinate system $(x,y,z)$, while the wider SMBHB is described in the C.O.M. coordinate system $(X,Y,Z)$.}
\label{fig:Binary SMBH scheme}
\end{figure}

In the presence of an SMBHB, the binary motion also contributes to the geometric delay periodically (see Fig.~\ref{fig:Binary SMBH scheme}). 
In contrast to the intrinsic apsidal precession, the outer orbital motion introduces an \emph{in-phase} modulation of the recurrence times of even and odd eruptions~\citep{Miniutti_2025}. 
However, Fig.~\ref{fig:GSN Padj} shows no hint of an \emph{in-phase} modulation except the \emph{anti-phase} modulation.
In the remainder of this section, we proceed to conduct a rigorous Bayesian inference and quantitatively constrain the excluded parameter space of the assumed companion SMBH.

Compared with the forced EMRI + equatorial disk hypothesis (base hypothesis hereafter) in~\cite{Zhou_2025}, 
we consider the binary-modulation hypothesis with the configuration shown in Fig.~\ref{fig:Binary SMBH scheme}, wherein the  tertiary SMBH with mass $M_e$ (denoted as $M_{e}$ hereafter) is in an assumed circular orbit around the central SMBH (mass $M_{\bullet}$, also denoted as $M_{\bullet}$ hereafter) of the EMRI. The separation between the two SMBHs is $a_{\rm out}$. The distance $A_{\rm CM}$ from $M_\bullet$ to the SMBHB C.O.M. is therefore expressed as
\begin{equation}
    A_{\rm CM}=[M_e/(M_\bullet+M_e)]a_{\rm out}=a_{\rm out}/(1+q),
\end{equation}
where $q$ is the mass ratio $M_\bullet/M_e$. The angular velocity $\omega$ of the SMBHB is
\begin{equation}
    \omega=\sqrt{\frac{M_{\bullet}+M_{e}}{a_{\rm out}^3}}=\sqrt{\frac{M_{\bullet}}{a_{\rm out}^3}\left(1+\frac{1}{q}\right)}.
    \label{eq:SMBHB angular velocity}
\end{equation}
In the SMBHB plane in the C.O.M. coordinate $(X,\ Y,\ Z)$, the position vector of $M_\bullet$ is
\begin{equation}
    \boldsymbol{R}_{M_{\bullet}}=[A_{\rm CM}\cos\phi(t),\ A_{\rm CM}\sin\phi(t),\ 0],
\end{equation}
where $\phi(t)$ is the orbital phase of $M_\bullet$ in the SMBHB plane, measured with respect to C.O.M., with
\begin{equation}
\phi(t)=\omega\ t+\phi_{M_{\bullet}\ \rm ini}. 
\end{equation}
We denote the line-of-sight unit vector from the SMBHB C.O.M. to the observer by $\hat{\boldsymbol{n}}_{\rm obs,CM}$, specified by the angles $(\theta_{\rm obs,CM},\ \phi_{\rm obs,CM})$. The orbital motion of $M_\bullet$ around the C.O.M. induces an additional LTT modulation,
\begin{equation}
\begin{aligned}
    \delta t_{\rm binary}&=-\boldsymbol{R}_{M_{\bullet}}\cdot\hat{\boldsymbol{n}}_{\rm obs,CM}\\
    &=-A_{\rm CM}\sin\theta_{\rm obs,CM}\cos[\phi(t)-\phi_{\rm obs,CM}]\\
    &=-\mathcal{A}_{0}\cos(\omega\ t+\Delta\Phi_{0}),
\end{aligned}
\label{eq:delta_t}
\end{equation}
where $\mathcal{A}_{0}\equiv A_{\rm CM}\sin\theta_{\rm obs,CM}$ is the projected modulation amplitude, the modulation frequency $\omega$ is equivalent to the SMBHB angular velocity~\eqref{eq:SMBHB angular velocity}, and $\Delta\Phi_{0}$ is the initial phase offset between the SMBH $M_\bullet$ and the projected line of sight in the SMBHB orbital plane, $\Delta\Phi_{0}=\phi_{M_{\bullet}\ \rm ini}-\phi_{\rm obs,CM}$. 
An in-phase modulation to the arrival times of even and odd eruptions as in Eq.~(\ref{eq:delta_t}) could be a signature of an SMBHB.

We further define the observer direction in the EMRI coordinate system $(x,y,z)$ via the effective angles $(\theta_{\rm obs,eff},\ \phi_{\rm obs,eff})$, which arise from the SMBHB-induced transformation. Here $\theta_{\rm obs,eff}$ is equivalent to $\theta_{\rm obs}$ as adopted in~\citet{Zhou_2025}.
Without loss of generality, we set $\phi_{\rm obs,eff}=0$ by aligning the $x$ axis with the projection of the observer direction onto the EMRI equatorial plane. The unit vector pointing from the SMBH $M_{\bullet}$ to the SMO at the disk-crossing point is
\begin{equation}
    \hat{\boldsymbol{n}}_{\rm crs}=(\sin\theta_{\rm crs}\cos\phi_{\rm crs},\ \sin\theta_{\rm crs}\sin\phi_{\rm crs},\ \cos\theta_{\rm crs}),
\end{equation}
the projection of $\hat{\boldsymbol{n}}_{\rm crs}$ along the observer direction is then
\begin{equation}
    \hat{\boldsymbol{n}}_{\rm obs,eff}\cdot\hat{\boldsymbol{n}}_{\rm crs}=\sin\theta_{\rm crs}\sin\theta_{\rm obs,eff}\cos\phi_{\rm crs}+\cos\theta_{\rm crs}\cos\theta_{\rm obs,eff}.
\end{equation}
Therefore, the tertiary SMBH $M_{e}$ introduces three additional timing-modulation parameters, $\{\mathcal{A}_{0},\ \omega,\ \Delta\Phi_{0}\}$, to be constrained. 

With the QPE timing model and the GSN 069 data, one can infer the model parameters and compare different hypotheses by measuring
the Bayesian evidence. The likelihood definition, model parameter priors and Bayesian inference setup can be found in previous studies \citep{PhysRevD.109.103031,PhysRevD.110.083019,10.1093/mnras/staf1580,Zhou_2025} and are also briefly summarized in Appendix~\ref{appendix:fit GSN 069}.
The constraints of model parameters are presented in Fig.~\ref{fig:corner plot GSN069},
where constraints of $p,\ e,\ T_{\rm obt},\ \dot{T}_{\rm obt},$ and $M_{\bullet}$ are consistent with those obtained in~\cite{Zhou_2025}, specifically 
\begin{equation}
\begin{aligned}
e &= 0.04\pm 0.02\ , \\
    \log_{10}\left(M_{\bullet}/M_{\odot}\right) &= 5.6\pm 0.1\ , 
\end{aligned}
\end{equation}
at $95\%$ credible level. The EMRI apsidal precession naturally explains the anti-phase branches in $T_{\rm rec}(t)$ shown in Fig.~\ref{fig:GSN Padj}.
The posterior of $\mathcal{A}_0$ is consistent with zero with a $95\%$ credible upper limit
\begin{equation}
    \mathcal{A}_0 < 219 \,M_\bullet \ ,
    \label{eq:GSN069_A0_upper_limit_cong_prior}
\end{equation}
i.e., there is no evidence for an {in-phase} modulation caused by a companion SMBH.
The modulation frequency $\omega$ and phase $\Delta\Phi_{0}$ remain weakly constrained. 
This null detection result is also consistent with $T_{\rm rec}(t)$ as shown in Fig.~\ref{fig:GSN Padj} where no evidence for an in-phase modulation is revealed.

Consistent with the parameter constraints, we obtain the log Bayes factor between the base and binary-modulation hypotheses
\begin{equation}
    \log\mathcal{B}^{\rm binary}_{\rm base}=-3.15^{+0.26}_{-0.26}\ ,
    \label{eq:GSN069_binary_bayes_factor_cong_prior}
\end{equation}
which represents strong support for the base hypothesis, i.e., a companion SMBH hypothesis is strongly disfavored, according to Jeffreys’ scale.
The posterior corner plots of all model parameters with the binary-modulation hypothesis are shown in Fig.~\ref{fig:corner plot GSN069} in Appendix~\ref{appendix:fit GSN 069}.
\section{Indirect probe with survival of stars}\label{sec:indirect probe}
The survival of the SMO provides an additional, indirect constraint on the presence of the tertiary SMBH in the EMRI system. 
We consider a hierarchical triple system composed of an inner orbit and an outer perturbing object. As shown in Fig.~\ref{fig:Binary SMBH scheme}, the tertiary SMBH can form such a system with the inner EMRI provided that the outer orbital scale is sufficiently larger than that of the EMRI. 
The orbits are fully described by the Keplerian elements: the semi-major axis $a_{\rm orb}$, eccentricity $e_{\rm orb}$, inclination $i_{\rm orb}$ (relative to the reference plane); argument of pericenter $\omega_{\rm orb}$, longitude of ascending node $\Omega_{\rm orb}$, and true anomaly $f_{\rm orb}$. The first three define the orbital geometry, and the last three specify the phase. Unless otherwise stated, we replace the index ``orb" with ``in" for the inner EMRI orbit, ``out" for the outer orbit of the tertiary SMBH $M_e$ hereafter.

The ZLK mechanism can exchange inclination and eccentricity between the two orbits and drive the inner orbit to very high eccentricity~\citep{1962AJ.....67..591K,LIDOV1962719,annurev:/content/journals/10.1146/annurev-astro-081915-023315}. Such eccentricity excitation reduces the pericenter distance,
\begin{equation}
    r_{\rm p}=a_{\rm in}(1-e_{\rm in}).
\end{equation}

Several QPE scenarios invoke stellar EMRIs, in which the secondary is a normal star rather than a compact object such as a stellar-mass black hole (sBH)~\citep{Xian_2021,10.1093/mnras/stad2203,Linial_2023}; the survival of such a star
therefore requires its pericenter to remain outside the TDE radius $r_{\rm TDE}$.
For an EMRI with a central SMBH of mass $M_{\bullet}$, a star of mass $m_{\star}$, and radius $R_{\star}$ would be tidally disrupted once~\citep{10.1111/j.1365-2966.2005.08843.x}
\begin{equation}
    r_{\rm p}\lesssim r_{\rm TDE}\simeq R_{\star}\left(\frac{M_\bullet}{m_{\star}}\right)^{1/3}.
    \label{eq:TDE condition}
\end{equation}
This diagnostic applies only when the SMO is a normal star. If the SMO is a compact object, such as an sBH, the absence of a stellar TDE provides no constraint on the outer SMBH.
Therefore, if the EMRI has a surviving stellar orbiter, the absence of TDE over the observed lifetime requires the maximum eccentricity excited by the tertiary object to satisfy
\begin{equation}
    a_{\rm in}(1-e_{\rm in,max})\gtrsim r_{\rm TDE}.
\end{equation}
This survival condition excludes regions of the tertiary SMBH parameter space that would otherwise drive the SMO into the TDE regime. In practice, the limiting eccentricity should be evaluated including octupole-order perturbations and short-range forces such as relativistic apsidal precession, which can substantially modify the extreme-eccentricity excitation in hierarchical triples~\citep{10.1093/mnras/stu2396,Huang_2026}.

To quantify this survival constraint, we define $X\equiv (M_\bullet/M_{e})(a_{\rm out}/1500 M_\bullet)^3$, 
which is proportional to the GR precession
parameter~\citep{Huang_2026}
\begin{equation}
\begin{aligned}
    \varepsilon_{\rm GR} &\equiv\frac{3M_{\bullet}^2a_{\rm out}^3(1-e_{\rm out}^2)^{\frac{3}{2}}}{a_{\rm in}^4 M_{e}}\ , \\
    &= 1.25 (1-e_{\rm out}^2)^{\frac{3}{2}}  \left( \frac{M_\bullet}{M_e}\right)
    \left( \frac{a_{\rm out}}{1500 M_\bullet}\right)^{3} \left( \frac{a_{\rm in}}{300 M_\bullet}\right)^{-4}  \ ,
\end{aligned}
\end{equation}
and perform a direct scan in the $(i_{\rm in,\rm ini},X)$ plane for parameters motivated by GSN 069, based on the system depicted in Sec.~\ref{sec:modulation}. We adopt fiducial values $e_{\rm out}=0$ and $a_{\rm in}=300\,M_\bullet$.
Here, $a_{\rm out}$ and $e_{\rm out}$ are the semi-major axis and eccentricity of the relative outer orbit between the tertiary SMBH $M_{e}$ and the C.O.M. of the inner EMRI. In the test-particle (TP) limit and circular-orbit approximation, $a_{\rm out}$ is the SMBHB separation in Sec.~\ref{sec:modulation}.

We adopt the TP limit for the inner companion in computing the orbital dynamics, while the stellar mass and radius entering the TDE radius $r_{\rm TDE}$ are assumed as $m_\star=1\,M_\odot$ and $R_\star=1\,R_\odot$.
The inner orbit is characterized by the semi-major axis $a_{\rm in}=300\ M_\bullet$ around a central SMBH of mass $M_\bullet=4\times10^5\ M_\odot$.
The outer orbit is taken to be circular with $e_{\rm out}=0$, and the evolution is computed with the SA equations~\citep{Liu_2018} including octupole-order terms and 1PN relativistic
apsidal precession. This should be distinguished from the double-averaged limit: the conventional double-averaged octupole parameter is proportional to $e_{\rm out}$ as~\citep{Huang_2026}
\begin{equation}
    \varepsilon_{\rm Oct}=\frac{a_{\rm in}}{a_{\rm out}}\frac{e_{\rm out}}{1-e_{\rm out}^2},
\end{equation}
which vanishes for a circular outer orbit, whereas in the present SA treatment the instantaneous outer orbital phase is retained and the octupole-order contributions are not eliminated by the outer-orbit average. The evolution equations for the parameters of the hierarchical system in the TP limit are given in Eqs.~\eqref{eq:first SA eq}--\eqref{eq:last SA eq} in Appendix ~\ref{appendix:SA eq}.

For each prescribed value of $\varepsilon_{\rm GR}$, we realize the same relativistic precession strength with a logarithmic grid of tertiary SMBH masses, $10^5\leq M_{e}/M_\odot\leq10^6$, by solving for the corresponding outer semi-major axis $a_{\rm out}$. We impose the hierarchy cut $a_{\rm out}/a_{\rm in}\geq2.8$. The scan covers $20^\circ\leq i_{\rm in,ini}\leq160^\circ$, where $i_{\rm in}$ is the inclination angle of the inner orbit relative to the outer orbital plane in the TP limit, and $0.25\leq\varepsilon_{\rm GR}\leq5$, corresponding to
$0.2\leq X\leq4$.
We omit the exact value $i_{\rm in,\rm ini}=90^\circ$ from the discrete inclination grid, as the resulting extreme-eccentricity evolution is highly sensitive and numerically challenging to resolve accurately. For each grid point we also sample eight initial values of $\omega_{\rm in}$, while fixing $\Omega_{\rm in,ini}=0$, $f_{\rm out, ini}=0$, and $\omega_{\rm out}=180^\circ$. Each realization is integrated up to $\tau\equiv t/t_{\rm ZLK}=1000$, with $t_{\rm ZLK}$ the characteristic ZLK timescale,
\begin{equation}
    t_{\rm ZLK}=\sqrt{\frac{a_{\rm in}^3}{M_{\bullet}}}\frac{M_{\bullet}}{M_{e}}\left(\frac{a_{\rm out}}{a_{\rm in}}\right)^3(1-e_{\rm out}^2)^{\frac{3}{2}}.
\end{equation}
Then we record the maximum eccentricity $e_{\rm in, max}$ and the corresponding minimum pericenter distance $r_{\rm p,min}$.

\begin{figure*}[!ht]
\centering
\includegraphics[width=0.9\textwidth]{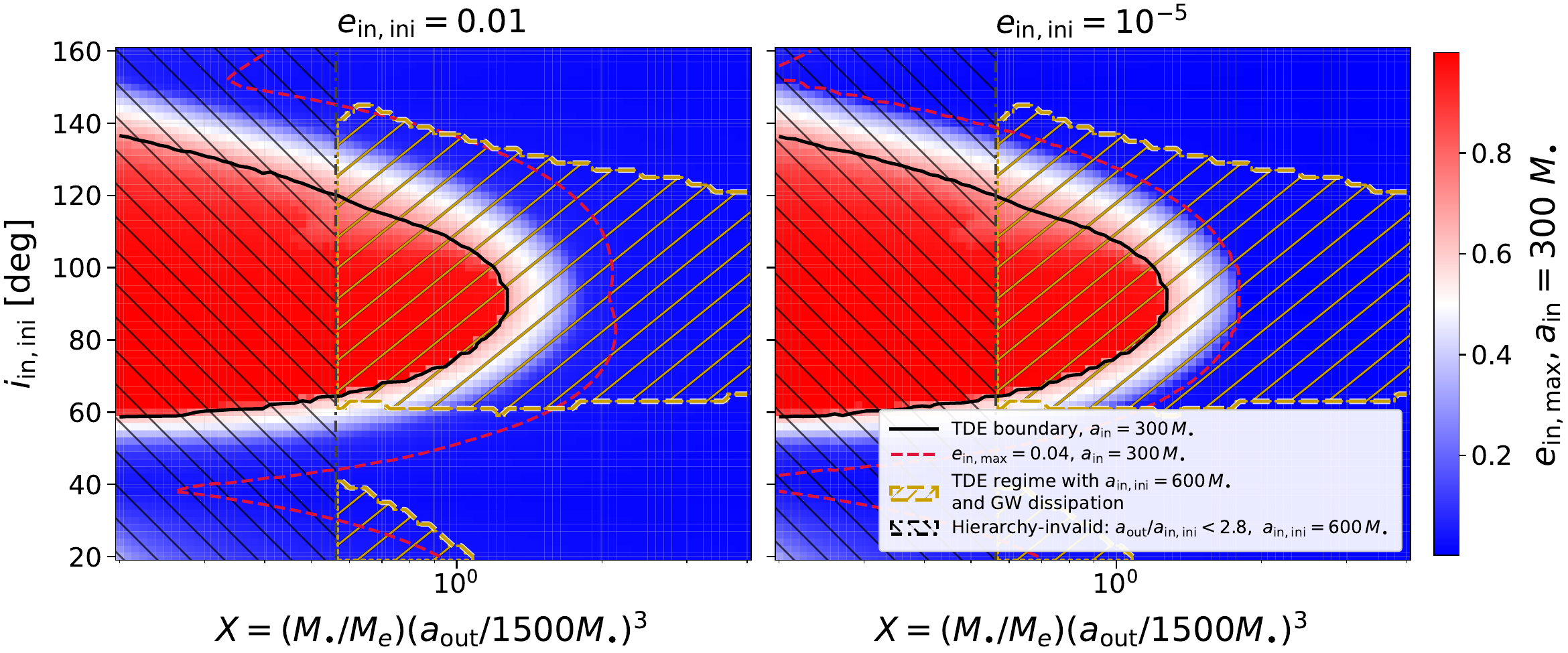}
\caption{{\bf Left:} Survival map of a star in the $(i_{\rm in,\rm ini},X)$ plane for a fiducial EMRI configuration motivated by the typical orbital scale of GSN 069. The color indicates the maximum inner orbital eccentricity $e_{\rm in,\rm max}$ reached during the SA evolution including octupole-order terms and 1PN relativistic apsidal precession. The calculation uses $M_\bullet=4\times10^5\,M_\odot$, $a_{\rm in}=300\,M_\bullet$, $e_{\rm in,ini}=0.01$, and a circular outer orbit with $e_{\rm out}=0$. 
For each displayed point, the plotted value is the largest $e_{\rm in,\rm max}$ obtained after sampling a logarithmic grid of tertiary SMBH masses $10^5\leq M_{e}/M_\odot\leq10^6$ and eight initial values of $\omega_{\rm in}$, subject to the hierarchy cut $a_{\rm out}/a_{\rm in}\geq2.8$. The black solid contour denotes the TDE boundary $a_{\rm in}(1-e_{\rm in,\rm max})=r_{\rm TDE}$, and the red dashed contour denotes $e_{\rm in,\rm max}=0.04$. Regions enclosed by the TDE boundary would tidally disrupt a solar-type star and are therefore disfavored if the observed QPE source contains a surviving star; {\bf Right:} Same as the left panel, but with a much lower initial inner orbital eccentricity $e_{\rm in,\rm ini}=10^{-5}$;
The yellow shaded regions in both panels denote the TDE regime obtained from a re-scan of the same $(M_{e},a_{\rm out})$ grid with $a_{\rm in,ini}=600\ M_\bullet$ and  GW dissipation.
} 
\label{fig:e1 distribution i1 epsilon GR e1 0.01 and 1e-5}
\end{figure*}

The resulting survival map is shown in Fig.~\ref{fig:e1 distribution i1 epsilon GR e1 0.01 and 1e-5}. 
The color represents the largest $e_{\rm in,\rm max}$ obtained among the allowed $M_{e}$ and phase realizations. The black solid contour marks the TDE boundary, $r_{\rm p,\rm min}=a_{\rm in}(1-e_{\rm in,\rm max})=r_{\rm TDE}$, while the red dashed contour marks $e_{\rm in,\rm max}=0.04$. The high-eccentricity region is concentrated at low $X$ and $i_{\rm in,ini}$ near $90^{\circ}$. For a set of fixed $a_{\rm in}$ and $e_{\rm out}$, increasing $X$ is equivalent to increasing $\varepsilon_{\rm GR}$, so relativistic apsidal precession increasingly suppresses the ZLK eccentricity excitation.
To examine the dependence of the survival map on the assumed $e_{\rm in,ini}$, we repeat the scan with nearly circular initial inner orbits for two initial values, with $e_{\rm in,\rm ini}=10^{-2}$ in the left panel, and $e_{\rm in,\rm ini}=10^{-5}$ in the right panel of Fig.~\ref{fig:e1 distribution i1 epsilon GR e1 0.01 and 1e-5}, while keeping all other physical parameters, phase sampling, hierarchy cuts, and integration settings unchanged. 
Therefore, if the SMO in GSN~069 is a surviving star, 
 the region enclosed by the TDE contour is excluded.

Compared with the fiducial $e_{\rm in,\rm ini}=0.01$ case, the $e_{\rm in,\rm ini}=10^{-5}$ initialization produces a larger low-excitation region below the diagnostic contour $e_{\rm in, max}=0.04$, while the overall distribution morphology remains similar.
In the numerical outputs, the number of displayed cells with $e_{\rm in,\rm max}\leq0.04$ increases from 2441 to 3086 out of 5600 cells. By contrast, the TDE regime is almost unchanged: the number of cells satisfying $r_{\rm p,\rm min}\leq r_{\rm TDE}$ changes only from 1299 to 1300. This indicates that the low-eccentricity contour is sensitive to the assumed initial eccentricity, whereas the TDE boundary is insensitive to the initial eccentricity.

In the analysis above, we have fixed the inner orbit at $a_{\rm in} = 300 M_\bullet$. In fact, the stellar EMRI should have migrated inward from a larger distance to the current location driven by GW emission. In the early stage of the migration, the higher value of $a_{\rm in}$ suppresses the GR precession and a larger region of SMBHB parameter space is expected to be excluded by
the survival of the stellar EMRI.
To illustrate this point, we include the GW dissipation in the orbital evolution of the stellar EMRI~\citep{Peters_1964},
\begin{equation}
\left.\dv{a_{\rm in}}{t}\right|_{\rm GW}
 =-\frac{64}{5}\frac{m_\star M_\bullet(M_\bullet+m_\star)}
 {a_{\rm in}^{3}(1-e_{\rm in}^{2})^{7/2}}
 \left(1+\frac{73}{24}e_{\rm in}^{2}
 +\frac{37}{96}e_{\rm in}^{4}\right),
\label{eq:peters coupled a}
\end{equation}
\begin{equation}
\left.\dv{e_{\rm in}}{t}\right|_{\rm GW}
 =-\frac{304}{15}\frac{m_\star M_\bullet(M_\bullet+m_\star)e_{\rm in}}
 {a_{\rm in}^{4}(1-e_{\rm in}^{2})^{5/2}}
 \left(1+\frac{121}{304}e_{\rm in}^{2}\right).
\label{eq:peters coupled e}
\end{equation}
We repeat the complete scan for both initial eccentricities with $a_{\rm in,ini}=600\ M_{\bullet}$, using the same $(i_{\rm in,ini}, X)$ grid and phase sampling as above. At each pixel, only realizations satisfying $a_{\rm out}/a_{\rm in,ini}\geq 2.8$ are retained. We evolve the system by coupling the ZLK equations in Appendix~\ref{appendix:SA eq} to the Peters GW terms~\eqref{eq:peters coupled a}--\eqref{eq:peters coupled e}, allowing $a_{\rm in}(t)$ and $e_{\rm in}(t)$ to evolve self-consistently.
The trajectories entering the TDE regime are terminated at the first disruption event, while surviving trajectories are evolved to $t=1000\ t_{\rm ZLK,ini}$.
The complete re-scan contains 3640 hierarchy-valid displayed pixels in each panel. 1863 and 1865 pixels enter the TDE regime for $e_{\rm in,ini}=0.01$ and $10^{-5}$, respectively; 1487 and 1490 of these pixels, respectively, are outside the TDE regime in the original $a_{\rm in}=300\ M_{\bullet}$ calculations. Across all 7280 hierarchy-valid pixels, the maximum $t_{\rm ZLK,ini}/t_{\rm GW}$ is $7.77\times 10^{-8}$,  where
$t_{\rm GW}\equiv\left|a_{\rm in}/\dot{a}_{\rm in}\right|_{\rm GW}$ is evaluated, for the realization defining each pixel, at its maximum-eccentricity state, or at the first TDE crossing for trajectories that undergo disruption;
 and the maximum fractional decrease in $a_{\rm in}$ over the integration, $\abs{\Delta a_{\rm in}}/a_{\rm in,ini}$, is $4.52\times 10^{-6}$. Thus, GW-driven migration is strongly subdominant to ZLK dynamics in determining the survival of the stellar orbiter.

This auxiliary calculation should therefore be interpreted as a migration survival test: if the same outer SMBH were already present while the stellar orbiter resided at the wider orbit, a larger region of parameter space would be vulnerable to TDE before GW emission appreciably shrank the inner orbit.

\section{Summary}\label{sec:summary}
In this paper, we investigated whether QPEs produced in an EMRI+disk system can be used to probe the SMBH companion. In this scenario, the central SMBH $M_{\bullet}$ in the EMRI and the tertiary SMBH $M_{e}$ form a wider SMBHB. We developed two complementary diagnostics: a direct constraint from the LTT modulation of the QPE arrival times, and an indirect constraint based on the survival of the star in the inner EMRI against tidal disruption driven by eccentricity excitation via the ZLK mechanism.

In Sec.~\ref{sec:review}, we first reviewed how the recurrence pattern of GSN~069 encodes the dynamics of the EMRI+disk system. In the disk-crossing interpretation, two eruptions are generally produced during each orbital period. A mildly eccentric orbit gives rise to alternating long and short recurrence intervals, $T_{\rm long}(t)$ and $T_{\rm short}(t)$~\citep{Miniutti2019}, which form two branches that vary approximately in anti-phase, while their sum remains close to the EMRI orbital period as shown in Fig.~\ref{fig:GSN Padj}. The relative variation amplitude in the two branches provides an estimate of the orbital eccentricity, whereas the modulation period of the two branches traces the relativistic apsidal-precession period. The combination of the orbital and apsidal-precession periods therefore provides a dynamical measurement of the central SMBH mass, i.e., the QPE timing method \citep{10.1093/mnras/staf1580,Zhou_2025}.

For the direct probe in Sec.~\ref{sec:modulation}, we incorporated the orbital motion of $M_{\bullet}$ around the SMBHB C.O.M. into the QPE timing model. This motion produces an additional LTT delay and, in contrast to the anti-phase modulation associated with apsidal precession, induces an in-phase modulation in the even and odd recurrence-time branches~\citep{Miniutti_2025}. The binary-modulation hypothesis introduces three additional parameters, $\{\mathcal{A}_{0},\omega,\Delta\Phi_{0}\}$, describing the modulation amplitude, orbital frequency, and initial phase difference. We fitted this model to the QPE arrival times of GSN~069 and found no evidence for an SMBHB-induced in-phase timing modulation, with $\mathcal{A}_{0}$ being consistent with zero.

For the indirect probe in Sec.~\ref{sec:indirect probe}, we treated the EMRI and the tertiary SMBH as a hierarchical triple system. If the SMO is a normal star, eccentricity excitation by the ZLK mechanism~\citep{1962AJ.....67..591K,LIDOV1962719,annurev:/content/journals/10.1146/annurev-astro-081915-023315,10.1093/mnras/stu2396} can reduce its pericenter below the TDE radius, so the survival of the star excludes companion configurations that enter the TDE regime. We computed survival maps in the $(i_{\rm in,ini},\ X)$ plane for a fiducial GSN~069-motivated configuration, with a circular outer orbit. The orbital evolution was calculated in the TP limit using the SA equations~\citep{Liu_2018} including octupole-order perturbations and 1PN relativistic apsidal precession~\citep{Huang_2026}. The high-eccentricity region is concentrated near $i_{\rm in,ini}\simeq90^{\circ}$, and at low $X$, corresponding to low $\varepsilon_{\rm GR}$. As $\varepsilon_{\rm GR}$ increases, relativistic apsidal precession increasingly suppresses the ZLK eccentricity excitation and enlarges the star-survival region. Repeating the calculation for $e_{\rm in,ini}=10^{-2}$ and $10^{-5}$ shows that the diagnostic contour $e_{\rm in,max}=0.04$, motivated by the fit results of GSN~069, depends appreciably on the assumed initial eccentricity. By contrast, the TDE boundary $a_{\rm in}(1-e_{\rm in,max})=r_{\rm TDE}$ remains nearly unchanged, indicating that the resulting survival constraint is considerably more robust.
A complete re-scan starting from $a_{\rm in,ini}=600\ M_{\bullet}$ enlarges the TDE regime, while the negligible fractional change in $a_{\rm in}$ confirms that ZLK dynamics, rather than GW-driven shrinkage, determines the TDE outcome over the integration.

\section*{Acknowledgements}
We thank Xiumin Huang for valuable discussions. The computations in this paper were run on the cluster supported by the Astronomy and Astrophysics Division of Tsung-Dao Lee Institute, Shanghai Jiao Tong University. \par
 Yuhe Zeng is supported by T.D. Lee scholarship.\par
Bin Liu acknowledges support from the National Natural Science Foundation of China (grant No. 12673087 and 12433008) and National Key Research and Development Program of China (No. 2023YFB3002502).

\textit{Software}: \texttt{bilby}~\citep{Ashton_2019}, \texttt{nessai}~\citep{PhysRevD.103.103006}.

\appendix

\section{Details of Bayesian inference}\label{appendix:fit GSN 069}

The likelihood $\mathcal{L}$ is constructed  from the observed QPE arrival times $t^{\rm obs}$ and model-predicted arrival times $t^{\rm model}_k$, where $k$ is the global cycle number. 
Due to large data gaps, there is uncertainty in assigning global cycle numbers to $t^{\rm obs}$ and we marginalize the uncertainty in the likelihood definition as follows.
Let $N_e$ denote the number of observed QPEs in epoch $e$, where $e\in\{\mathrm{XMM3, XMM4}, \mathrm{Chandra,XMM5}\}$, and the observed QPE arrival time is labeled locally as $t^{\rm obs}_{e, i}$,
where $0\leq i \leq N_e-1$.
The likelihood is defined as 
\begin{equation}
\begin{aligned}
    \mathcal{L}=\prod_{e} \mathcal{L}_e \ ,  \quad \mathcal{L}_e =\sum_{s_e} \prod_{i=0}^{N_e-1}\frac{1}{\sqrt{2\pi\Tilde{\sigma}^2_{i}}}\exp\left(-\frac{\abs{t_{e, i}^{\rm obs}-t_{s_e+i}^{\rm model}}^2}{2\Tilde{\sigma}^2_{i}}\right),  
    \label{eq:total likelihood}
\end{aligned}
\end{equation}
where $s_e$ is the global cycle number assigned to the first eruption in epoch $e$ and the sum $\sum_{s_e}$ collects contributions of all possible $s_e$. In practice, only a small number of $s_e$ dominate the contribution to the likelihood.
We define $\Tilde{\sigma}_{i}^{2}\equiv\sigma_{i}^2+\sigma_{\rm sys}^2$, with $\sigma_{i}$ and $\sigma_{\rm sys}$ denoting the measurement and unmodeled timing uncertainties, respectively.



The model parameter priors used in this work are summarized in
Table~\ref{tab:GSN_priors}.  The base hypothesis uses the identical shared
priors and likelihood construction but contains no
$\{\mathcal{A}_0,\omega,\Delta\Phi_0\}$ parameters.

\begin{table*}[!ht]
\centering
\caption{Priors for the GSN~069 binary-modulation timing fit.}
\label{tab:GSN_priors}
\begin{tabular}{llll}
\hline\hline
Parameter & Prior & Unit or boundary & Role\\
\hline
$p$                         & $\mathcal{U}(50,500)$       & $M_\bullet$              & Semilatus rectum\\
$e$                         & $\mathcal{U}(0,0.9)$        & ---                       & Eccentricity\\
$\cos\theta_{\min}$         & $\mathcal{U}(0,1)$          & ---                       & Polar turning point\\
$q_{r,\rm ini}$                   & $\mathcal{U}(0,2\pi)$       & periodic                  & Initial radial phase\\
$q_{z,\rm ini}$                   & $\mathcal{U}(0,2\pi)$       & periodic                  & Initial polar phase\\
$q_{\phi,\rm ini}$                & $\mathcal{U}(0,2\pi)$       & periodic                  & Initial azimuthal phase\\
$T_{\rm obt}$             & $\mathcal{U}(60,70)$        & ks                        & Initial orbital period\\
$a$                         & $\mathcal{U}(0,1)$          & ---                       & Dimensionless spin\\
$|\dot T_{\rm obt}|$        & $\mathcal{U}(0,10^{-4})$    & ---                       & Period-decay rate\\
$\theta_{\rm obs,eff}$      & $\mathcal{U}(0,\pi/2)$      & rad                       & Effective viewing angle\\
$\sigma_{\rm sys}$          & $\mathcal{U}(0,500)$        & ks                        & Additional timing scatter\\
$\mathcal{A}_0$              & $\mathcal{U}(0,10^5)$       & $M_\bullet$              & Projected LTT amplitude\\
$\omega$                    & $\mathcal{U}(0,10^{-4})$    & $M_\bullet^{-1}$         & Outer orbit angular frequency\\
$\Delta\Phi_0$              & $\mathcal{U}(0,2\pi)$       & periodic                  & Outer orbital phase\\
\hline
\end{tabular}
\end{table*}

\begin{figure*}[!ht]
\centering
\includegraphics[width=1\textwidth]{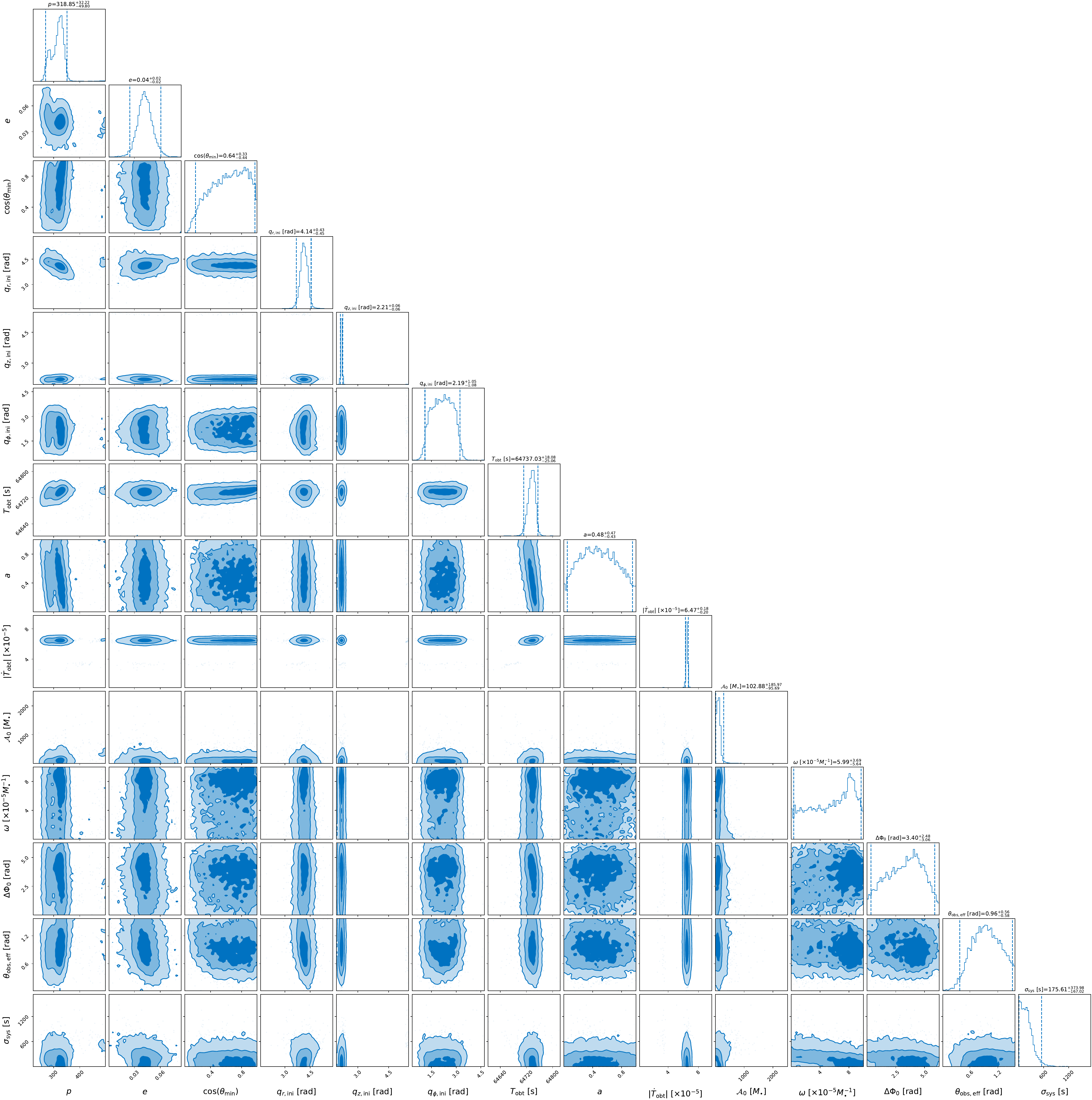}
\caption{Corner plot of the posterior distributions inferred from the GSN 069 QPE timing
data. The sampled parameters are
$p[M_{\bullet}],\ e,\ \cos\theta_{\rm min},\ q_{r,{\rm ini}},\ q_{z,{\rm ini}},\
q_{\phi,{\rm ini}},\ T_{\rm obt}[\rm sec],$ $a$, $\dot{T}_{\rm obt},\
\mathcal{A}_{0}[M_{\bullet}],\ \omega[M_{\bullet}^{-1}],\ \Delta\Phi_{0},\ \theta_{\rm obs,eff},\ \sigma_{\rm sys}[\rm sec]$.
The parameters $\mathcal{A}_{0}$, $\omega$, and $\Delta\Phi_{0}$ characterize
the SMBHB-induced time modulation.}
\label{fig:corner plot GSN069}
\end{figure*}

The model parameter inference is performed with the \texttt{nessai}~\citep{PhysRevD.103.103006} algorithm within \texttt{bilby}~\citep{Ashton_2019}, using 1000 live points. Sampling is terminated when the estimated remaining change in the log evidence satisfies $\Delta\ln\mathcal{Z}\leq 0.1$.
The posterior distribution corner plots for GSN 069 with the binary-modulation hypothesis are shown in Fig.~\ref{fig:corner plot GSN069}.
The inference reveals an eccentric EMRI and no evidence for an SMBHB, which naturally explains the anti-phase modulation in even and odd eruptions shown in Fig.~\ref{fig:GSN Padj}.  

\section{Injection-recovery test of sensitivity to binary modulation}
\label{sec:injection_recovery}

In this section, we perform the injection-recovery experiments for testing the capability of the Bayesian inference in extracting in-phase modulations in QPE timing, using the same four-epoch sampling structure as the GSN~069 analysis.

We first generate mock QPE timing data similar to those of GSN 069, except with an underlying circular EMRI in an SMBHB.
The model parameters used for generating mock data are summarized in Table~\ref{tab:injection_values}, where the binary-modulation amplitude $\mathcal{A}_{0}$ and frequency $\omega$ are consistent with \citet{Miniutti_2025}, 
\begin{equation}
    \mathcal{A}_{0}=9306\ {\rm s}=2.585\ {\rm hr},\qquad \omega=3.7935\times10^{-6}\ {\rm rad\,s^{-1}}.
\end{equation}
This frequency corresponds to the period $P_{\rm binary}=19.17\ {\rm d}$. The mock timing data used for the injection-recovery test are summarized in Table~\ref{tab:mock data injection recovery}.

\begin{table*}[!ht]
\centering
\caption{Injected parameters for the injection-recovery test.}
\label{tab:injection_values}
\begin{tabular}{llll}
\hline\hline
Parameter & Injected value & Parameter & Injected value\\
\hline
$p$ & $300\,M_\bullet$ & $e$ & $0$\\
$\cos\theta_{\min}$ & $0.8$ & $q_{r,\rm ini}$ & $0$\\
$q_{z,\rm ini}$ & $5.3\ {\rm rad}$ & $q_{\phi,\rm ini}$ & $2.0\ {\rm rad}$\\
$T_{\rm obt}$ & $64800\ {\rm s}$ & $a$ & $0.5$\\
$|\dot T_{\rm obt}|$ & $5\times10^{-5}$ & $\theta_{\rm obs,eff}$ & $0.4\ {\rm rad}$\\
$\mathcal{A}_0$ & $9306\ {\rm s}$ & $\omega$ & $3.7935\times10^{-6}\ {\rm rad\,s^{-1}}$\\
$\Delta\Phi_0$ & $0$ & $\sigma_{\rm sys}$ & $75\ {\rm s}$\\
\hline
\end{tabular}
\end{table*}

\begin{figure*}[!ht]
\centering
\includegraphics[width=\textwidth]{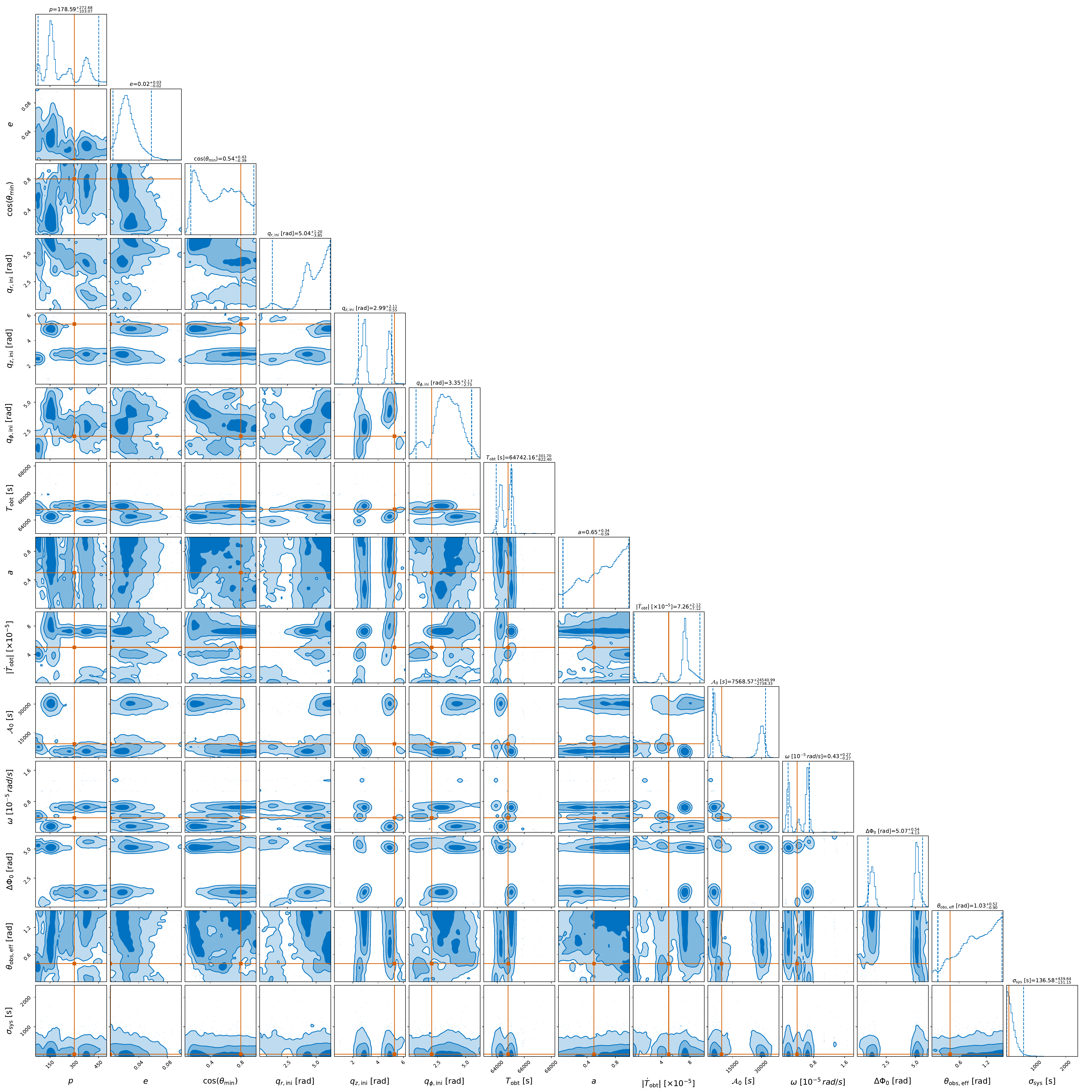}
\caption{Posterior distributions for the binary-modulation injection-recovery test.  Orange lines and points indicate the injected values; the injected $q_{r,\rm ini}$ is not marked because the radial phase is unidentifiable for an exactly circular orbit, and $\Delta\Phi_0=0$ is shown at the equivalent $2\pi$ boundary.}
\label{fig:injection_recovery_corner}
\end{figure*}

We perform the same Bayesian inference of the mock data, and the posteriors obtained are shown in Fig.~\ref{fig:injection_recovery_corner}. 
The injection values of model parameters are correctly recovered, with EMRI orbital eccentricity $e < 0.05$ (95\% credible upper limit), and binary modulation parameters 
\begin{equation}
    \mathcal{A}_0=7568.6^{+24541.0}_{-2738.3}\ {\rm s},\quad \omega=\left(4.265^{+2.683}_{-2.747}\right)\times10^{-6}\ {\rm rad\,s^{-1}} ,
\end{equation} 
where the corresponding binary period is constrained as $P_{\rm binary}=17.051^{+30.861}_{-6.584}\ {\rm d}$.
Here the central values denote posterior medians and the quoted uncertainties correspond to the central 95\% credible intervals. 
Different from the case of real GSN 069 data, the EMRI orbital size $p$ is not constrained in this experiment, simply because there is no anti-phase modulation in the mock data,
and therefore the apsidal precession period $T_{\rm aps}$ is not constrained.

\begin{table*}[!ht]
\centering
\caption{The mock timing data used for the injection-recovery test. The fixed reference time is
$t_{\rm ref}=662030428.5588458~{\rm s}$;
the listed $\sigma_i$ are the measurement uncertainties.}
\label{tab:mock data injection recovery}
\begin{tabular}{lrr}
\hline\hline
Epoch & $t_i^{\rm mock}-t_{\rm ref}$ [s] & $\sigma_i$ [s]\\
\hline
XMM3    &     -8907.452531 & 114.000000\\
XMM3    &     23017.453261 &  83.000000\\
\hline
XMM4    &   1971949.031710 & 397.000000\\
XMM4    &   2005593.605607 & 180.000000\\
XMM4    &   2038657.171337 &  97.000000\\
XMM4    &   2072727.832073 & 163.000000\\
XMM4    &   2105607.493478 &  74.000000\\
\hline
Chandra &   4530213.180895 & 378.538710\\
Chandra &   4560650.840953 & 408.786830\\
Chandra &   4592709.921364 & 487.773990\\
\hline
XMM5    &  13700835.738031 & 358.000000\\
XMM5    &  13734182.607940 &  95.000000\\
XMM5    &  13766646.928871 &  86.000000\\
XMM5    &  13800455.774488 & 162.000000\\
XMM5    &  13832774.749017 & 191.000000\\
\hline
\end{tabular}

\medskip
\end{table*}

\begin{table*}[!ht]
\centering
\caption{Injection-recovery results for eccentric EMRI orbits, with $p=300\ M_{\bullet}$ and $e=0.04$.} 
\label{tab:eccentric_injection_results}
\begingroup
\small
\setlength{\tabcolsep}{4pt}
\begin{tabular}{lccccc}
\hline\hline
Run & $\mathcal A_{0,\rm inj}$ [ks] & $P_{\rm binary,inj}$ [d] & $\log\mathcal B^{\rm binary}_{\rm base}$
& $\mathcal A_0$ [ks] posterior & $P_{\rm binary}$ [d] posterior\\
\hline
T0  & 0     & ---   & $-3.75\pm0.28$ & $5.229^{+5.154}_{-5.201}$ & ---\\
T1  & 4.680 & 19.17 & $+1.67\pm0.25$ & $10.461^{+1.555}_{-9.261}$ & $22.781^{+93.201}_{-19.380}$\\
T2 & 9.306 & 19.17 & $+4.66\pm0.25$ & $5.230^{+13.432}_{-1.819}$  & $7.969^{+11.963}_{-3.294}$\\
T3  & 9.306 & 160   & $+2.86\pm0.25$ & $4.870^{+10.526}_{-1.933}$  & $106.984^{+146.928}_{-94.666}$\\
T4 & 9.306 & 320   & $-1.67\pm0.25$ & $3.348^{+9.515}_{-3.291}$  & $27.714^{+6.462}_{-26.381}$\\
\hline
\end{tabular}
\endgroup
\vspace{2pt}
\begin{minipage}{\textwidth}
\footnotesize
\end{minipage}
\end{table*}

We extend the above test to eccentric EMRI orbits with $p=300\,M_\bullet$ and $e=0.04$, keeping the other intrinsic parameters in Table~\ref{tab:injection_values} unchanged. We perform five injection-recovery tests, including one without binary modulation and four with nonzero modulation, with different modulation amplitudes and periods. For each injection, we
perform the inference with the two different hypotheses and calculate the log Bayes factor, $\log\mathcal B^{\rm binary}_{\rm base}$.
The results are summarized in Table~\ref{tab:eccentric_injection_results}.

In T0, where no binary modulation was injected, the Bayes factor strongly disfavors the binary hypothesis, similar to GSN 069.
In T1-3, where a binary modulation with a short period $P_{\rm binary,inj} \leq 160$ d was injected, the Bayes factor strongly favors the binary hypothesis. In T4, where a binary modulation with a long period $P_{\rm binary,inj} = 320$ d was injected, the Bayes factor strongly disfavors the binary hypothesis. Due to the short observation span $\approx 160$ d, it is of no surprise to find the long-period modulation injected is not resolved. These tests illustrate the sensitivity limit of QPE observations to the binary modulation.

\section{O-C consistency check for GSN 069}\label{appendix:O-C GSN 069}
In this section, we present O-C analysis for GSN~069 as an independent diagnostic (see \cite{zhou2026noteqpetimingfalse} for a complete analysis). We first construct the O-C values relative to a fixed linear ephemeris,
\begin{equation}
    \Delta t_{\rm O-C}^{\rm lin}\equiv t_{\rm obs}-t_{\rm ref}-T_{\rm rec,trial}N_{\rm cyc},
\end{equation}
where $t_{\rm ref}$ is the reference epoch, chosen as the first observed QPE, $T_{\rm rec,trial}=8.94\,{\rm hr}$ is a fixed trial recurrence time, and $N_{\rm cyc}$ is the global QPE cycle number.

We fit these O-C values with a quadratic secular baseline plus a parity-dependent sinusoidal modulation. Equivalently, the full toy timing model is
\begin{equation}
    t_{\rm toy}(N_{\rm cyc})
    =t_{\rm ref}+\delta t
    +T_{\rm rec}N_{\rm cyc}
    -\frac{1}{2}|\dot{T}_{\rm rec}|\bar{T}_{\rm rec}
    N_{\rm cyc}^2 +A_{\rm toy}\sin\left(\frac{2\pi N_{\rm cyc}}{P_{\rm toy}}+\phi_s\right),
    \label{eq:toy model}
\end{equation}
where $\delta t$, $T_{\rm rec}$, and $|\dot{T}_{\rm rec}|$ are phenomenological parameters of the toy model, with
\begin{equation}
    \bar{T}_{\rm rec}
    =
    T_{\rm rec}
    -\frac{1}{2}|\dot{T}_{\rm rec}|\Delta t_{\rm span},
\end{equation}
with $\Delta t_{\rm span}$ denoting the total observational time span. The sinusoidal component has a common amplitude $A_{\rm toy}$ and period $P_{\rm toy}$ for the two timing branches, while $\phi_s$ is a parity-dependent phase, with $s\in\{\mathrm{even},\mathrm{odd}\}$.
We denote the quadratic baseline and sinusoidal component by
\begin{equation}
    B_{\rm quad}\equiv \delta t+(T_{\rm rec}-T_{\rm rec,trial})N_{\rm cyc}-\frac{1}{2}|\dot{T}_{\rm rec}|\bar{T}_{\rm rec}N_{\rm cyc}^2
\end{equation}
and
\begin{equation}
    S_s\equiv A_{\rm toy}\sin\left(\frac{2\pi N_{\rm cyc}}{P_{\rm toy}}+\phi_s\right),
\end{equation}
respectively.

\begin{figure}[htbp]
    \centering
    \subfigure{\includegraphics[width=0.45\textwidth]{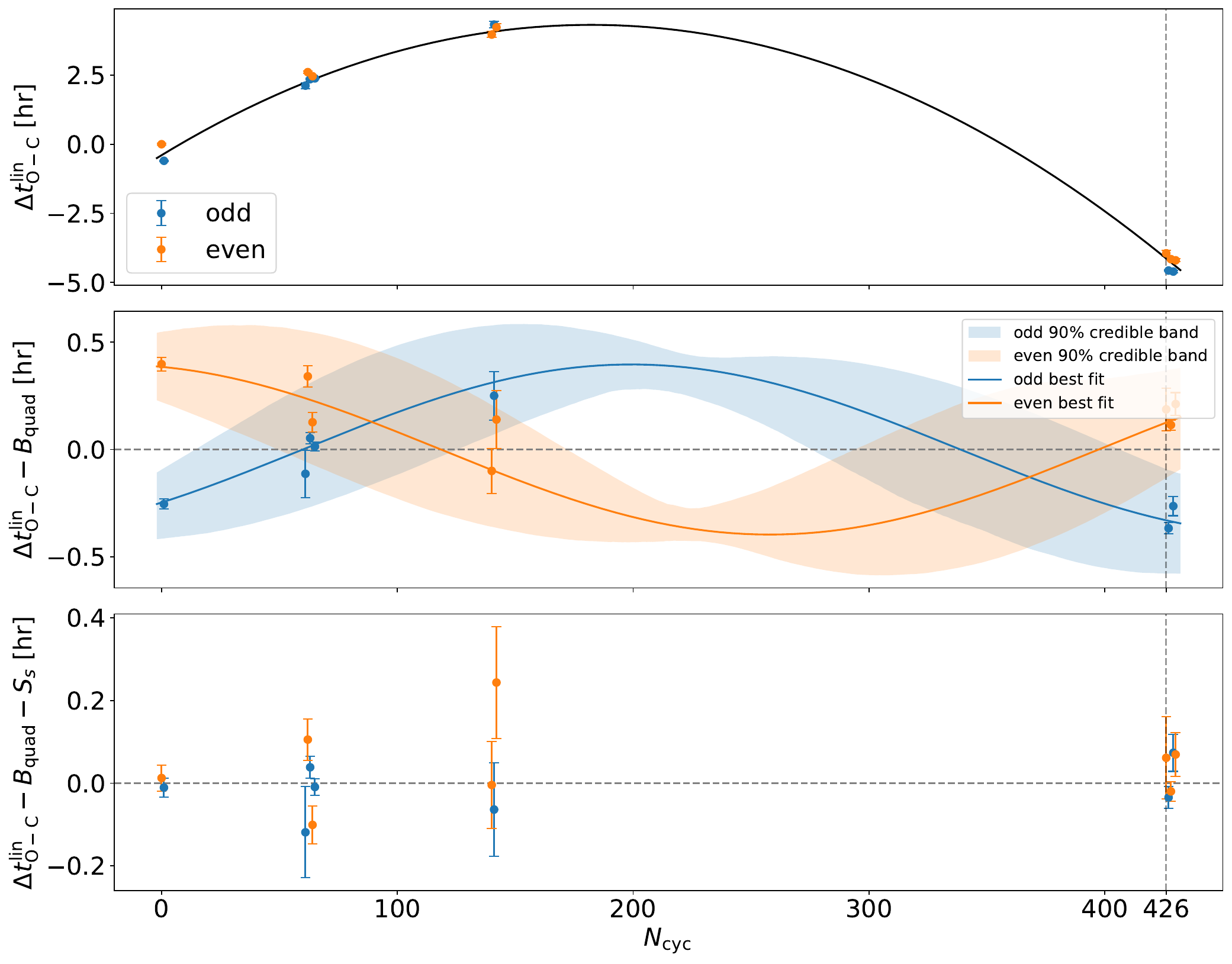}}
    \subfigure{\includegraphics[width=0.45\textwidth]{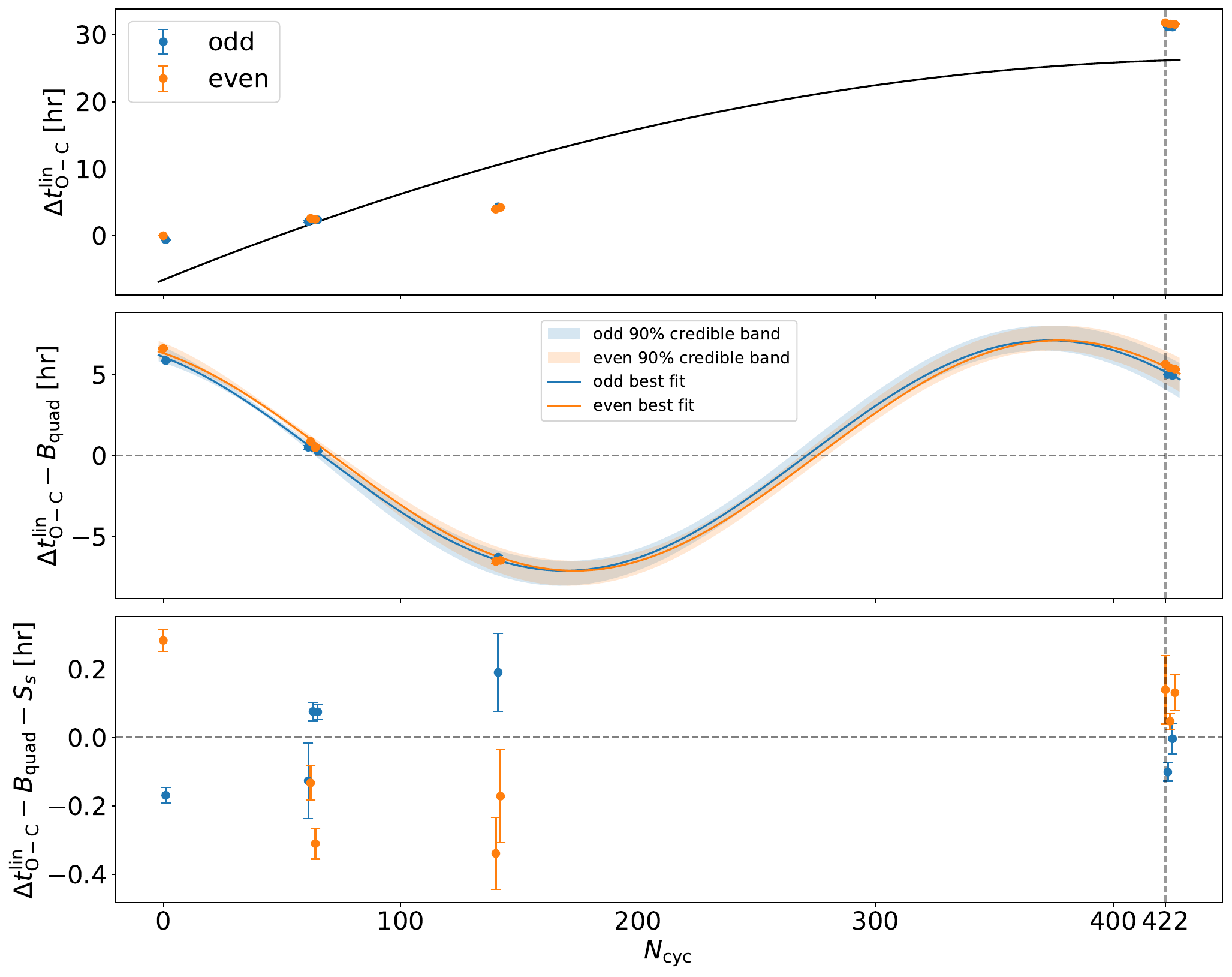}}
    \caption{O-C analysis of GSN~069 under two global cycle assignments.
Left: Reference assignment with $N_{\rm FinalEpochStart}=426$.
From top to bottom, the panels show the O-C values relative to the fixed linear ephemeris, the baseline-subtracted modulation
$\Delta t_{\rm O-C}^{\rm lin}-B_{\rm quad}$, and the residuals after further subtracting the parity-dependent sinusoidal component,
$\Delta t_{\rm O-C}^{\rm lin}-B_{\rm quad}-S_s$.
The middle panel shows the best-fit odd and even modulations together with their 90\% credible bands, which reveals two anti-phase branches.
Right: Same as the left, but for the shifted assignment $N_{\rm FinalEpochStart}=422$, which produces an apparent in-phase modulation between the odd and even branches.}
    \label{fig:O-C fitting}
\end{figure}

As the toy model~\eqref{eq:toy model} depends explicitly on the cycle number $N_{\rm cyc}$,  a mismatch in $N_{\rm cyc}$ leads to false signals in the O-C diagram. As shown by \cite{zhou2026noteqpetimingfalse}, a universal feature of these false signals is a large \emph{in-phase} modulation in even and odd eruptions.
We denote by $N_{\rm FinalEpochStart}$ the global $N_{\rm cyc}$ assigned to the first QPE in the final observing epoch (XMM5). We take $N_{\rm FinalEpochStart}=426$ as the reference assignment and compare it with the shifted assignment $N_{\rm FinalEpochStart}=422$. As shown in Fig.~\ref{fig:O-C fitting}, the reference assignment yields an anti-phase modulation of the odd and even residuals, whereas the shifted assignment produces an in-phase modulation.

From the two O-C fits alone, it may not be straightforward to identify which analysis is correct due to the dependence on the unknown cycle number $N_{\rm cyc}$.
On the contrary, there is no such ambiguity in the recurrence time $T_{\rm rec}(t)$ (see Fig.~\ref{fig:GSN Padj}), which clearly reveals the anti-phase branches 
without requiring a global cycle count across observational gaps. Therefore, an O-C fit that fails to recover the anti-phase modulation is incorrect.

\section{Equations of evolution}\label{appendix:SA eq}
The evolution equations for the inner orbital elements following the formulation of~\citet{Liu_2018}, including the SA octupole-order perturbation and relativistic apsidal precession, are shown in Eqs.~\eqref{eq:first SA eq}--\eqref{eq:last SA eq}. For simplicity, we denote the inner orbit with index ``$1$", and the outer orbit with index ``$2$", corresponding to ``in" and ``out" in Sec.~\ref{sec:indirect probe}, respectively. 
\begin{equation}
    \left.\dv{e_1}{\tau}\right|_{\rm Quad}=5\nu_{\tau} e_1j_1AB,
    \label{eq:first SA eq}
\end{equation}
\begin{equation}
    \left.\dv{i_1}{\tau}\right|_{\rm Quad}=\nu_{\tau} Q\left[j_1 B\sin\omega_1+\frac{1+4e_1^2}{j_1}A\cos\omega_1\right],
\end{equation}
\begin{equation}
    \left.\dv{\Omega_1}{\tau}\right|_{\rm Quad}=\nu_{\tau}\sin(\Omega_1-u_2)\left(\frac{1+4e_1^2}{j_1}A\sin\omega_1-j_1 B\cos\omega_1\right),
\end{equation}
\begin{equation}
    \begin{aligned}
        \left.\dv{\omega_1}{\tau}\right|_{\rm Quad}&=\nu_{\tau} j_1(Q^2+5A^2-2)-\dv{\Omega_1}{\tau}\Big|_{\rm Quad}\cos i_1\\
        &=\nu_{\tau} j_1(Q^2+5A^2-2)-\nu_{\tau} \cos i_1\sin(\Omega_1-u_2)\left(\frac{1+4e_1^2}{j_1}A\sin\omega_1-j_1 B\cos\omega_1\right).
    \end{aligned}
\end{equation}
\begin{equation}
    \left.\dv{e_1}{\tau}\right|_{\rm Oct}=-\nu_{{\rm Oct},\tau}j_1B\mathcal{M}_{\rm Oct}.
\end{equation}
\begin{equation}
    \left.\dv{i_1}{\tau}\right|_{\rm Oct}=-\nu_{{\rm Oct},\tau}e_1Q\left[10j_1AB\sin\omega_1+\frac{10j_1^2A^2+\mathcal{M}_{\rm Oct}}{j_1}\cos\omega_1\right].
\end{equation}
\begin{equation}
\begin{aligned}
    \left.\dv{\Omega_1}{\tau}\right|_{\rm Oct}&=\frac{\nu_{{\rm Oct},\tau}e_1Q}{\sin i_1}\left[10j_1AB\cos\omega_1-\frac{10j_1^2A^2+\mathcal{M}_{\rm Oct}}{j_1}\sin\omega_1\right]\\
    &=\nu_{{\rm Oct},\tau}e_1\sin(\Omega_1-u_2)\left[10j_1AB\cos\omega_1-\frac{10j_1^2A^2+\mathcal{M}_{\rm Oct}}{j_1}\sin\omega_1\right].
\end{aligned}
\end{equation}
\begin{equation}
    \left.\dv{\omega_1}{\tau}\right|_{\rm Oct}=\nu_{{\rm Oct},\tau}\frac{j_1A}{e_1}\left(16e_1^2-\mathcal{M}_{\rm Oct}-10e_1^2Q^2\right)-\left.\dv{\Omega_1}{\tau}\right|_{\rm Oct}\cos i_1.
\end{equation}
\begin{equation}
    \left.\dv{\omega_1}{\tau}\right|_{\rm GR}=\frac{\varepsilon_{\rm GR}}{1-e_1^2}.
\end{equation}
where
\begin{equation}
    \tau\equiv \frac{t}{t_{\rm ZLK}},
\end{equation}
\begin{equation}
    \nu_{\tau}=\frac{3}{2}\frac{(1+e_2\cos f_2)^3}{(1-e_2^2)^{\frac{3}{2}}},
\end{equation}
\begin{equation}
    \nu_{{\rm Oct},\tau}=\frac{15}{16}\frac{a_1}{a_2}\frac{(1+e_2\cos f_2)^4}{(1-e_2^2)^{5/2}},
\end{equation}
and 
\begin{equation}
    \dv{f_2}{\tau}=t_{\rm ZLK}\sqrt{\frac{M_{\bullet}+M_{e}}{a_{2}^{3}}}\frac{(1+e_2\cos f_2)^2}{(1-e_2^2)^{\frac{3}{2}}}.
\end{equation}
Defining
\begin{equation}
    u_2\equiv\omega_2+f_2,
\end{equation}
we have
\begin{equation}
    A=\cos\omega_{1}\cos(\Omega_1-u_2)-\sin\omega_1\cos i_1\sin(\Omega_1-u_2),
    \label{eq: A}
\end{equation}
\begin{equation}
    B=\sin\omega_1\cos(\Omega_1-u_2)+\cos i_1\cos\omega_1\sin(\Omega_1-u_2),
    \label{eq:B}
\end{equation}
\begin{equation}
    Q=\sin i_1\sin(\Omega_1-u_2).
    \label{eq: Q}
\end{equation}
There are also
\begin{equation}
    j_1\equiv \sqrt{1-e_1^2},
\end{equation}
\begin{equation}
    \mathcal{M}_{\rm Oct}\equiv 1-8e_1^2+35e_1^2A^2-5j_1^2Q^2.
\end{equation}
Finally,
\begin{equation}
    \dv{e_1}{\tau}=\left.\dv{e_1}{\tau}\right|_{\rm Quad}+\left.\dv{e_1}{\tau}\right|_{\rm Oct},
\end{equation}
\begin{equation}
    \dv{i_1}{\tau}=\left.\dv{i_1}{\tau}\right|_{\rm Quad}+\left.\dv{i_1}{\tau}\right|_{\rm Oct},
\end{equation}
\begin{equation}
    \dv{\Omega_1}{\tau}=\left.\dv{\Omega_1}{\tau}\right|_{\rm Quad}+\left.\dv{\Omega_1}{\tau}\right|_{\rm Oct},
\end{equation}
\begin{equation}
    \dv{\omega_1}{\tau}=\left.\dv{\omega_1}{\tau}\right|_{\rm Quad}+\left.\dv{\omega_1}{\tau}\right|_{\rm Oct}+\left.\dv{\omega_1}{\tau}\right|_{\rm GR}.
    \label{eq:last SA eq}
\end{equation}
\section{Tidal Effects on the star}
\begin{figure*}[!ht]
\centering
\includegraphics[width=0.5\textwidth]{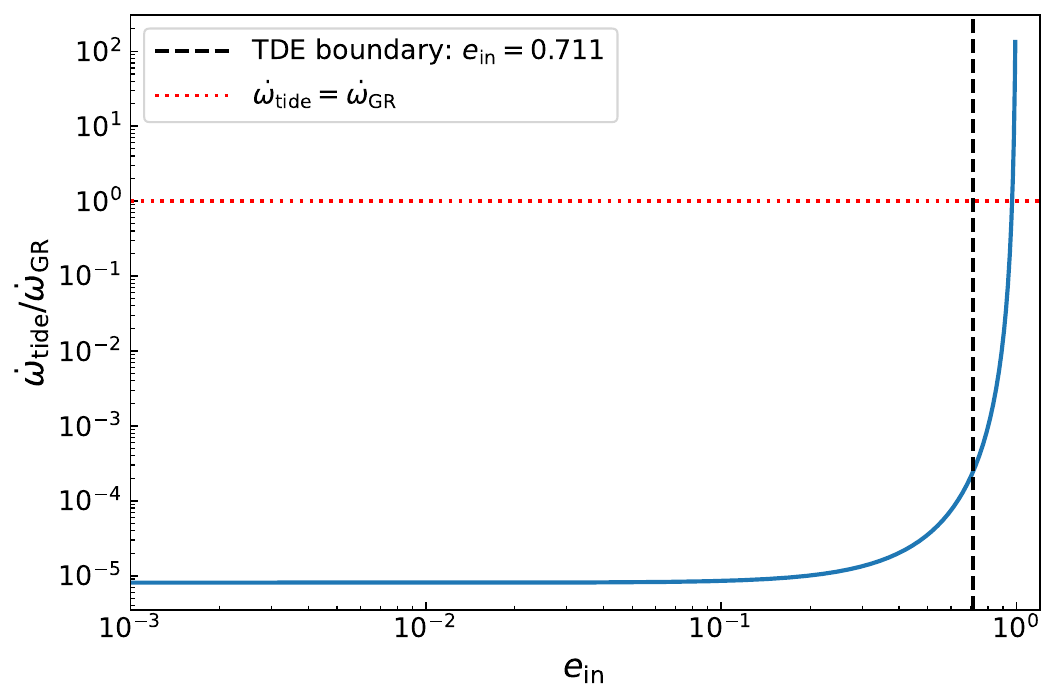}
\caption{Comparison of $\dot{\omega}_{\rm tide}$ and $\dot{\omega}_{\rm GR}$ as a function of $e_{\rm in}$. The vertical dashed line marks $e_{\rm in}=0.711$, which corresponds to the TDE boundary for $a_{\rm in}=300\ M_{\bullet}$ in this work. The horizontal red dotted line indicates $\dot{\omega}_{\rm tide}=\dot{\omega}_{\rm GR}$.} 
\label{fig:omega tide omega GR}
\end{figure*}

A normal star can be tidally distorted by the central SMBH.  In addition to the TDE condition~\eqref{eq:TDE condition} used in Sec.~\ref{sec:indirect probe}, the
equilibrium tidal bulge produces an additional apsidal precession of the inner orbit.  Following \citet{10.1093/mnras/stu2396}, its contribution in the TP limit can be expressed as
\begin{equation}
    \left.\dv{\omega_{\rm in}}{\tau}\right|_{\rm tide}
    =\varepsilon_{\rm tide}
    \frac{1+\frac{3}{2}e_{\rm in}^{2}
    +\frac{1}{8}e_{\rm in}^{4}}
    {(1-e_{\rm in}^{2})^{5}},
\end{equation}
where
\begin{equation}
    \varepsilon_{\rm tide}
    =15k_{2,\star}\frac{M_{\bullet}^{2}}{m_{\star}M_{e}}
    \frac{a_{\rm out}^{3}}{a_{\rm in}^{8}}
    (1-e_{\rm out}^{2})^{3/2}R_{\star}^{5}.
\end{equation}
Here $k_{2,\star}$ is the stellar tidal Love number.  The stellar mass $m_{\star}$ is retained only in this physical tidal coefficient, whereas the orbital evolution remains in the strict TP limit.
For the fiducial parameters in this work, $m_{\star}=1\,M_{\odot}$,
$R_{\star}=1\,R_{\odot}$, and $k_{2,\star}=0.0144$, together with
$M_{\bullet}=4\times10^{5}\,M_{\odot}$ and
$a_{\rm in}=300\,M_{\bullet}$, the coefficient ratio is
$\varepsilon_{\rm tide}/\varepsilon_{\rm GR}=8.06\times10^{-6}$.  Including the dependence on eccentricity $e_{\rm in}$, as shown in Fig.~\ref{fig:omega tide omega GR}, the tidal apsidal rate is only
$2.40\times10^{-4}$ of the relativistic rate at the disruption boundary
\begin{equation}
    e_{\rm in}=1-r_{\rm TDE}/a_{\rm in}=0.711.
\end{equation}
The two rates become comparable only at $e_{\rm in}\simeq0.966$, corresponding to $r_{\rm p}\simeq0.118\,r_{\rm TDE}$, which lies deep within the TDE regime.


\bibliographystyle{aasjournal}
\bibliography{reference}
\end{document}